\documentclass[twocolumn,nofootinbib,amsmath,amssymb,aps,prd]{revtex4-1}

\usepackage[english]{babel}
\usepackage{amsmath,amssymb,amsfonts,latexsym,cancel}

\usepackage{mathtools}
\usepackage{graphicx}
\usepackage{color}
\usepackage{amsbsy}
\usepackage{hyperref}
\usepackage{tikz}

\newcommand{\be}{\begin{equation}}
\newcommand{\ee}{\end{equation}}

\newcommand{\bma}{\begin{bmatrix}}
\newcommand{\ema}{\end{bmatrix}}

\newcommand{\pd}{\partial}

\def\ea{{E}^{a}}
\def\eb{{E}^{b}}
\def\f{\frac}
\def\nn{\nonumber}
\def\lp{\ell_{\rm Pl}}

\begin{document}

\title{Black hole collapse and bounce in effective loop quantum gravity}

\author{Jarod George Kelly} \email{jarod.kelly@unb.ca}
\affiliation{Department of Mathematics and Statistics, University of New Brunswick, %\\
Fredericton, NB, Canada E3B 5A3}

\author{Robert Santacruz} \email{robert.santacruz@unb.ca}
\affiliation{Department of Mathematics and Statistics, University of New Brunswick, %\\
Fredericton, NB, Canada E3B 5A3}

\author{Edward Wilson-Ewing} \email{edward.wilson-ewing@unb.ca}
\affiliation{Department of Mathematics and Statistics, University of New Brunswick, %\\
Fredericton, NB, Canada E3B 5A3}

\begin{abstract}

We derive effective equations with loop quantum gravity corrections for the Lema\^itre-Tolman-Bondi family of space-times, and use these to study quantum gravity effects in the Oppenheimer-Snyder collapse model.  For this model, after the formation of a black hole with an apparent horizon, quantum gravity effects become important in the space-time region where the energy density and space-time curvature scalars become comparable to the Planck scale.  These quantum gravity effects first stop the collapse of the dust matter field when its energy density reaches the Planck scale, and then cause the dust field to begin slowly expanding.  Due to this continued expansion, the matter field will eventually extend beyond the apparent horizon, at which point the horizon disappears and there is no longer a black hole.  There are no singularities anywhere in this space-time.  In addition, in the limit that edge effects are neglected, we show that the dynamics for the interior of the star of uniform energy density follow the loop quantum cosmology effective Friedman equation for the spatially flat Friedman-Lema\^itre-Robertson-Walker space-time.  Finally, we estimate the lifetime of the black hole, as measured by a distant observer, to be $\sim (GM)^2/\ell_{\rm Pl}$.

\end{abstract}

\maketitle

\section{Introduction}

General relativity is typically expected to break down when the curvature reaches the Planck scale, at which point a theory of quantum gravity becomes necessary.  One important example is a black hole of mass $M$: in the simplest case of the Schwarzschild vacuum spherically symmetric space-time, the Kretschmann scalar $R_{\mu\nu\rho\sigma} R^{\mu\nu\rho\sigma} \sim (2GM)^2 / r^6$ reaches the Planck scale at the radius $r \sim (2 GM \lp^2)^{1/3}$, suggesting that quantum gravity effects may become important not only close to the singularity at $r=0$, but as far away from the black hole center as $(2 GM \lp^2)^{1/3}$.

In addition to quantum gravity effects being presumably important in vacuum black hole solutions, quantum gravity is also expected to play an important role during black hole collapse.  While the matter forming the star is classically predicted to reach the central singularity, quantum gravity effects could resolve the singularity and modify the dynamics of infalling matter.

In spherically symmetric space-times, there are no gravitational waves and therefore to study black hole collapse a matter field is needed.  A simple choice is pressureless dust; spherically symmetric space-times minimally coupled to dust are known as the Lema\^itre-Tolman-Bondi (LTB) solutions \cite{Lemaitre:1933gd, Tolman:1934za, Bondi:1947fta}.

To explore how quantum gravity effects may modify black hole collapse, we will consider the effect of expressing the (classical) Hamiltonian in terms of area and holonomy variables, as suggested by loop quantum gravity (LQG) \cite{Thiemann:2007pyv}.  (For the quantization of LTB space-times in other contexts, see \cite{Vaz:2000zb, Kiefer:2005tw, Kiefer:2019csi}.)  In this approach, the Hamiltonian depends on holonomies of the Ashtekar-Barbero connection along paths of physical length $\sim \lp$, rather than the connection itself.  The resulting `effective' equations of motion include quantum gravity effects due to the presence of these Planck-length holonomies in the Hamiltonian.  For cosmology, the effective equations derived in this manner have been shown to give an excellent approximation to the leading order quantum gravity effects for sharply-peaked states in loop quantum cosmology (LQC) \cite{Ashtekar:2011ni}.  Note that to ensure the physical length of the holonomy paths is $\sim \lp$, the coordinate length has to be related to the physical length through the metric; this step is essential to obtain physical results (otherwise an unphysical coordinate length is related to the physical Planck scale, resulting in inconsistencies) and this procedure is called, for historical reasons, the `improved dynamics' or `$\bar\mu$-scheme' \cite{Ashtekar:2006wn}.

There has been considerable work studying various LQG effects in spherically symmetric space-times.  For vacuum space-times, most studies are based on the isometry between the classical Schwarzschild interior and the Kantowski-Sachs space-time \cite{Modesto:2004xx, Ashtekar:2005qt, Bohmer:2007wi, Campiglia:2007pb, Chiou:2008nm, Brannlund:2008iw, Joe:2014tca, Corichi:2015xia, Cortez:2017alh, Olmedo:2017lvt, BenAchour:2018khr, Ashtekar:2018lag, Bodendorfer:2019cyv, Alesci:2019pbs, Assanioussi:2019twp}, but this isometry is based on results in classical general relativity and may not hold in LQG (it is also unclear how to properly carry out the $\bar\mu$ scheme when a spatial coordinate becomes null at a horizon).  The full vacuum space-time (interior and exterior) was first studied without using the $\bar\mu$ scheme \cite{Bojowald:2005cb, Gambini:2008dy, Reyes:2009, Gambini:2013ooa, Gambini:2013hna, BenAchour:2016brs, Bojowald:2018xxu}, and more recent work has shown how to include that missing step and implement the $\bar\mu$ scheme in vacuum spherically symmetric space-times \cite{Chiou:2012pg, Gambini:2020nsf, Kelly:2020uwj}.

To study black hole collapse, it is necessary to include a matter field.  For previous work on the inclusion of matter fields in spherical symmetry in the context of LQG, see \cite{Gambini:2009ie, Reyes:2009, Gambini:2014qta, Bojowald:2015zha, Campiglia:2016fzp}, and for studies of black hole collapse based on this approach, see \cite{Husain:2006cx, Husain:2008tc, Hossenfelder:2009fc, Campiglia:2016fzp, Benitez:2020szx}.  (And for other LQG/LQC-based studies of black hole collapse, see \cite{Ashtekar:2008jd, Ashtekar:2010qz, Tavakoli:2013rna, Bambi:2013caa, Liu:2014kra, Christodoulou:2016vny, Christodoulou:2018ryl, BenAchour:2020gon, Bianchi:2018}).  Although these studies provide interesting insights into potential quantum gravity effects in black hole collapse---in particular, many predict that the black hole singularity is resolved and that the black hole bounces into a white-hole-like solution---none use the $\bar\mu$ scheme throughout the entire space-time.  (Some of these studies do not use the $\bar\mu$ scheme at all, while others use it only for the interior of the star and use the equations of general relativity for the vacuum region outside the star.)   Finally, there have also been some studies of LQG-motivated inverse triad effects in spherical symmetry, for details see \cite{Husain:2004yz, Ziprick:2009nd, Bojowald:2009ih, Kreienbuehl:2010vc, Bojowald:2011js}.

In this paper we will show how to implement the $\bar\mu$ scheme for the full family of LTB space-times and then use this framework to derive effective equations of motion. These equations will hold both in the presence and absence of matter, as well as inside and outside a horizon.  We then use the resulting effective equations to study LQG effects in the Oppenheimer-Snyder black hole collapse model.  Note that we will refer to a black hole being present if there is, at that instant of (coordinate) time, an apparent horizon; we do not require an event horizon to exist since we do not necessarily expect an event horizon to be present in a non-singular black hole space-time.

\section{Classical Theory}
\label{s.class}

The line element for a spherically symmetric space-time can be put in the form
\be
ds^2 = -N^2 dt^2 + \f{(\eb)^2}{\ea} (dx+N^x dt)^2 + \ea d\Omega^2,
\ee
where $N(x,t)$ and $N^x(x,t)$ are the lapse and the radial component of the shift vector, $\ea(x,t)$ and $\eb(x,t)$ are the densitized triad in the radial and angular directions, and $d\Omega^2 = d\theta^2 + \sin^2\theta ~ d\phi^2$.

The densitized triads are conjugate to the Ashtekar-Barbero connection whose components are \cite{Kelly:2020uwj}
\begin{align}
A^i_a \tau_i dx^a = & \,\, a \, \tau_1 \, dx + \left( b \, \tau_2 + \f{\partial_x \ea}{2 \eb} \, \tau_3 \right) d\theta \nn \\
& \!\!\! + \left( \cot\theta \, \tau_1 - \f{\partial_x \ea}{2 \eb} \tau_2  + b \, \tau_3 \right) \sin\theta \, d\phi.
\end{align}
Here $a(x,t)$ and $b(x,t)$ capture the extrinsic curvature in the radial and angular directions respectively, while $\tau^j = -i \sigma^j / 2$, with $\sigma^j$ the Pauli matrices.

For the Lema\^itre-Tolman-Bondi (LTB) space-times the matter content is a pressureless dust field and, after integrating over $d\Omega$, the action is \cite{Husain:2011tk}
\begin{align}
S = \! \int \! dt \! \int \! dx \Biggr[ & \frac{\dot{a}\ea+2\dot{b}\eb}{2G\gamma} + 4\pi \dot{T} p_T
- N \left( \mathcal{H}^{(g)} + \mathcal{H}^{(d)} \right) \nn \\
& - N^x \left(\mathcal{H}^{(g)}_x  - 4 \pi p_T \pd_x T \right) \Biggr].
\end{align}
Here $T$ denotes the dust field and $p_{T}$ is its conjugate momentum, while the dots denote derivatives with respect to $t$.  The contributions to the scalar constraint from the gravitational and dust sectors are
\begin{align}
\mathcal{H}^{(g)} = &-\frac{1}{2 G \gamma^2}\biggr(2ab\sqrt{\ea}+\frac{\eb}{\sqrt{\ea}}(b^2+\gamma^2)\biggr) \nn \\
& +\frac{1}{8 G}\frac{(\partial_x \ea)^2}{\eb\sqrt{\ea}} + \f{\sqrt{\ea}}{2 G} \partial_x \left( \frac{\partial_x \ea}{\eb} \right), \\
%\end{align}
%
\mathcal{H}^{(d)} = & 4\pi \sqrt{p_T^2+\frac{\ea}{(\eb)^2}p_T^2(\pd_xT)^2},
\end{align}
and the gravitational term in the diffeomorphism constraint is
$\mathcal{H}^{(g)}_x = (2 G \gamma)^{-1} \cdot (2 \eb \pd_x b - a \pd_x \ea )$.

The Hamiltonian framework can be simplified by using the dust-time gauge to fix the scalar constraint \cite{Husain:2011tk} and the areal gauge to fix the diffeomorphism constraint \cite{Campiglia:2007pb, Kelly:2020uwj}.  After this gauge-fixing there will remain a true Hamiltonian, with no constraints left.  Note that due to this gauge-fixing procedure, the resulting effective theory does not fall within the class of (vacuum) models studied in \cite{Aruga:2019dwq}.

The benefit of the dust-time gauge $T=t$ is clear, as then $\pd_x T = 0$.  The gauge-fixing condition $\chi_1 = T-t = 0$ is second-class with the scalar constraint, so $\chi_1$ can be used to gauge-fix it.  Solving the scalar constraint gives $4 \pi p_T = -\mathcal{H}^{(g)}$, while requiring that the gauge-fixing condition be preserved by the dynamics imposes $N=1$.  Further, the symplectic term in the action $4 \pi p_T \dot{T}$ simplifies to $4 \pi p_T = -\mathcal{H}^{(g)}$, which becomes a true physical Hamiltonian $\mathcal{H}_{\rm phys} = \mathcal{H}^{(g)}$ \cite{Husain:2011tk}.

The next simplification is to impose the areal gauge through the condition $\chi_2 = \ea - x^2 = 0$, which imposes that a sphere at radius $x$ has surface area $4\pi x^2$.  The condition $\chi_2$ is second-class with the diffeomorphism constraint, which can be solved giving $a = \eb (\pd_x b) / x$, and requiring that $\chi_2$ be preserved dynamically gives $N^x = -b/\gamma$ \cite{Kelly:2020uwj}.  Since $\ea$ is independent of time, the $\dot a \ea$ term is a total time derivative and can be dropped from the action, while $\mathcal{H}_{\rm phys}$ simplifies considerably after substituting for $\ea$ and $a$:
\begin{align}
S_{GF} =& \int dt \int dx \left( \f{\dot{b}\eb}{G\gamma} - \mathcal{H}_{\rm phys} \right), \\
\mathcal{H}_{\rm phys} =& - \, \frac{1}{2G\gamma} \biggr[ \frac{\eb}{\gamma x} \left( b^2 + x\pd_x b^2 \right) + \frac{\gamma \eb}{x} \nn \\ & \qquad \qquad
+ \frac{2 \gamma x^2}{(\eb)^2} \pd_x \eb - \frac{3 \gamma x}{\eb} \biggr].
\end{align}

After fixing these two gauges, the metric reduces to
\be \label{ltb-pg}
ds^2 = - dt^2 + \f{(\eb)^2}{x^2} \left( dx + N^x \, dt \right)^2 + x^2 d\Omega^2,
\ee
with $N^x = - b/\gamma$ for classical general relativity, and it is clear that there is one physical degree of freedom at each point due to the dust field (there are no gravitational waves in spherically symmetric space-times).  The energy density $\rho(x,t)$ of the dust field, related to the dust contribution to the scalar constraint by $\mathcal{H}^{(d)} = \int d\Omega \, \sqrt{q} \rho$, where $\sqrt{q}$ is the determinant of the spatial metric, is
\be \label{density}
\rho = \f{p_T}{\sqrt{\ea} \, \eb} = - ~ \f{\mathcal{H}_{\rm phys}}{4 \pi x \, \eb},
\ee
and the remaining non-trivial Poisson bracket is
\be \label{pb}
\{b(x_1,t), \eb(x_2,t)\} = G \gamma \, \delta(x_1-x_2).
\ee
The dynamics follow from $\dot f = \{f, \int \! dx ~ \mathcal{H}_{\rm phys} \}$, giving
\begin{align}
\dot \eb & = \f{b \eb}{\gamma x} - \f{b}{\gamma} \, \pd_x \eb, \\
\dot b &= \f{\gamma x}{2 (\eb)^2} - \f{1}{2 \gamma x} \Big( 2x b \pd_x b + b^2 + \gamma^2 \Big).
\end{align}
These are the usual equations of motion for the LTB family of metrics, for the Painlev\'e-Gullstrand coordinates \eqref{ltb-pg} of the metric \cite{Lasky:2006hq, Giesel:2009jp}, and in a form convenient to include holonomy corrections as motivated by LQG.

\section{LTB Effective Equations}

One of the main features of LQG is that the fundamental operators, out of which all other operators are constructed, are holonomies of $A_a$ and the areas $E^a$.  In LQC, it has been shown that there exist `effective equations' that provide a good approximation to the dynamics of expectation values of observables, at least for wave functions that are sharply peaked and whose expectation value for the spatial volume satisfies $\langle V \rangle \gg \lp^3$ \cite{Taveras:2008ke, Rovelli:2013zaa}.  These effective equations can be derived as the Hamilton equations of an effective Hamiltonian that includes modifications proportional to $\hbar$ which ensure the resulting dynamics track sharply-peaked wave functions in the full quantum theory.

To derive an effective Hamiltonian for (gauge-fixed) LTB space-times, the main step is to replace the connection component $b$ by minimal length holonomies of $b$ in the classical gauge-fixed Hamiltonian.  (Recall that after the gauge-fixing imposed in Sec.~\ref{s.class}, the system is described by a single true Hamiltonian---there are no constraints left.) It is necessary that these holonomies give trigonometric functions of $b$ (without $b$-dependent prefactors) for it to be possible to promote the holonomies to operators in LQC.  This condition is satisfied by holonomies of the extrinsic curvature 1-form in the $\theta$ direction,
\begin{align}
h_\theta(2 \delta_b) &= \exp \left( \int_0^{2\delta_b} \! b \tau_2 \, d\theta \right) \nn \\ &
 = \cos \left( \delta_b b \right) \mathbb{I} + 2 \sin \left( \delta_b b \right) \tau_2.
\end{align}
This is known as the `K' loop quantization, for details see \cite{Vandersloot:2006ws, Singh:2013ava}.  (The path could be a portion of any great circle on the sphere, for simplicity we took $\phi = $ const.)

Still following \cite{Vandersloot:2006ws, Singh:2013ava}, the coordinate length $2 \delta_b$ must be chosen so that the physical length of the path is $\sqrt\Delta$, where $\Delta \sim \lp^2$ is the smallest non-zero eigenvalue of the area operator in LQG.  (The factor of 2 is to ensure consistency with expressions of the curvature in terms of holonomies \cite{Ashtekar:2009um}.)  For this path, $x$ and $\phi$ are constant, so the metric \eqref{ltb-pg} implies $ds = x \, d\theta$.  Integrating and requiring that the physical length be $\sqrt\Delta$ gives $\int_0^{\sqrt\Delta} ds = \int_0^{2 \delta_b} x \, d\theta$, so $2 \delta_b = \sqrt\Delta / x$, see also \cite{Chiou:2012pg, Gambini:2020nsf, Kelly:2020uwj}.

Finally, $b$ must be replaced in $\mathcal{H}_{\rm phys}$ by an appropriate expression in terms of $h_\theta(2\delta_b)$ \cite{Vandersloot:2006ws, Singh:2013ava},
\be
b \rightarrow  \f{-2 \, {\rm Tr} (h_\theta(2\delta_b) \cdot \tau_2)}{2\delta_b} = 
\f{x}{\sqrt\Delta} \sin\left( \f{\sqrt\Delta}{x} b \right).
\ee
This substitution gives the effective physical Hamiltonian for LTB space-times,
\begin{align} \label{phys-ham-lqg}
\mathcal{H}_{\rm phys}^{LQG} = - \, \f{1}{2G\gamma} \Biggr[ &
\frac{\eb}{\gamma x} \pd_x \left( \frac{x^3}{\Delta} \, \sin^2 \! \frac{\sqrt{\Delta} \, b}{x} \right)
-\frac{3\gamma x}{\eb} \nn \\ &
+\frac{2\gamma x^2}{(\eb)^2}\pd_x \eb
+\frac{\gamma \eb}{x} \Biggr].
\end{align}
To reconstruct the space-time metric from the phase space variables it is necessary to find the shift $N^x$, and this requires rewriting the relation between $N^x$ and $b$ in terms of holonomies (as before, this step is required since there is no operator corresponding to $b$ in the quantum theory).  Following earlier work in vacuum spherically symmetric space-times gives \cite{Gambini:2020nsf, Kelly:2020uwj}
\be \label{shift-gen}
N^x = -\f{x}{\gamma \sqrt\Delta} \sin\frac{\sqrt{\Delta} \, b}{x} \cos\frac{\sqrt{\Delta} \, b}{x}.
\ee
Note that the form of the metric \eqref{ltb-pg} remains unchanged.

The effective dynamics for general LTB space-times follow directly from $\mathcal{H}_{\rm phys}^{LQG}$ and the (unchanged) Poisson bracket \eqref{pb},
\begin{align} \label{eq:ebdot}
\dot{\eb} =& - \frac{x^2}{\gamma \sqrt{\Delta}} \pd_x \! \left(\frac{\eb}{x}\right)
\sin\frac{\sqrt{\Delta} \, b}{x} \cos\frac{\sqrt{\Delta} \, b}{x}, \\
\label{eq:bdot}
\dot{b} = \frac{\gamma}{2} \Big(& \frac{x}{(\eb)^2}-\frac{1}{x}\Big)
-\f{1}{2\gamma \Delta x} \pd_x \left(x^3 \sin^2 \f{\sqrt\Delta \, b}{x} \right).
\end{align}
These effective equations can be used to study LQG effects in LTB space-times, and in particular in black hole collapse models.

The vacuum solutions $\rho=0$ correspond to $\mathcal{H}_{\rm phys}^{LQG} = 0$ and have already been studied \cite{Gambini:2020nsf, Kelly:2020uwj}.  In Painlev\'e-Gullstrand coordinates, the metric for the vacuum solution has the form
\begin{align} \label{metric}
ds^2=&-\left(1 - \f{R_S}{x} + \f{\gamma^2\Delta R_S^2}{x^4} \right)dt^2 + dx^2 \nn \\ &
+ 2\sqrt{\frac{R_S}{x}\left(1-\f{\gamma^2\Delta R_S}{x^3}\right)}dt\,dx
+ x^2d\Omega^2,
\end{align}
where $R_S = 2 G M$ is the Schwarzschild radius, and the quantum gravity corrections are proportional to $\Delta$.  Note that for the vacuum case $M=0$, the result is the classical Minkowski solution without any quantum gravity corrections.  Interestingly, this metric is only valid for $x \ge x_{\rm min} = (\gamma^2 \Delta R_S)^{1/3}$ \cite{Gambini:2020nsf, Kelly:2020uwj}.  This is not surprising, since in spherical symmetry there are no gravitational waves and therefore a central potential must be generated by some distribution of matter.  Since LQC effects are known to bound $\rho \lesssim \rho_{\rm Pl}$ by the Planck scale, this suggests that to have a source of mass $M$, there must be some matter content out to (assuming maximal density) $x \sim (M / \rho_{\rm Pl})^{1/3}$, in qualitative agreement with the bound $x \ge x_{\rm min}$ for the vacuum solution.  In the simple model for black hole collapse considered next, this expectation will be shown to be precisely correct.

This vacuum solution has an outer apparent horizon located at $x_{\rm outer} \sim R_S - {\gamma^2 \Delta}/{R_S}$ and an inner apparent horizon at $x_{\rm inner} \sim x_{\rm min} + (\gamma^4 \Delta^2/27 R_S)^{1/3}$ \cite{Kelly:2020uwj}; matter that lies inside the inner horizon can remain at rest, or bounce and start to expand as required for a transition from a black hole to a white hole \cite{BenAchour:2020gon}.  In the Planck regime, quantum gravity repulsive effects counteract the classical gravitational attractive force, and the outgoing expansion becomes positive again for $x < x_{\rm inner}$.

\section{Black Hole Collapse and Bounce}

A simple model for black hole collapse is the Oppenheimer-Snyder model \cite{Oppenheimer:1939ue}.  This space-time belongs to the family of LTB solutions, and corresponds to a `star' of radius $L(t)$, with vacuum outside.  The star is composed of pressureless dust, and it is further assumed that the star has a spatially constant density $\rho$ within its interior: $\rho(x,t) = \rho(t)$ if $x \le L(t)$ and $\rho(x,t) = 0$ for $x > L(t)$.  As the star collapses, $L(t)$ will decrease and $\rho(t)$ will increase.  To simplify the analysis, for now we neglect edge effects due to the discontinuity in $\rho$ at the surface $L$ of the star.  Note that the interior of the Oppenheimer-Snyder model is a Friedman-Lema\^itre-Robertson-Walker (FLRW) cosmology, while the exterior is vacuum.

For a spatially flat interior, we set $\eb = x$ which is clearly a solution of \eqref{eq:ebdot}.  Then \eqref{eq:bdot} and \eqref{density} simplify to
\begin{align} \label{bdot-os}
\dot{b} &= -\f{1}{2\gamma \Delta x} \, \pd_x \left(x^3 \sin^2 \f{\sqrt\Delta \, b}{x} \right), \\
\label{rho-b}
\rho &= \f{1}{8 \pi G \gamma^2 \Delta \, x^2} \, \pd_x \left( x^3 \, \sin^2 \! \frac{\sqrt{\Delta} \, b}{x} \right),
\end{align}
and conservation of energy implies that $M = 4 \pi \rho L^3 / 3$ is constant.  The relation \eqref{rho-b} is easily inverted, and for the collapsing Oppenheimer-Snyder model this gives
\be \label{sinb-os}
\sin \f{\sqrt\Delta b}{x} =
\begin{cases} - \sqrt{\rho(t)/\rho_c} \qquad & \mbox{if } x \le L(t), \\
- \sqrt{3M/4\pi \rho_c x^3} \quad & \mbox{if } x > L(t), \end{cases}
\ee
where the critical energy density is $\rho_c = 3 / (8 \pi G \gamma^2 \Delta)$.  Note that the overall minus sign is chosen so $N^x > 0$ and the star collapses (for $\cos (\sqrt\Delta b / x) \ge 0$).

Combining \eqref{bdot-os} and the interior solution \eqref{sinb-os} gives an equation for $\dot\rho$, and using $\rho = 3M / 4 \pi L^3$ this becomes
\be
\left(\frac{\dot{L}}{L}\right)^2 = \frac{8\pi G}{3}\rho\left(1-\frac{\rho}{\rho_c}\right),
\ee
exactly the LQC effective Friedmann equation for spatially flat FLRW space-times with scale factor $L$ \cite{Ashtekar:2006wn}.  This is easily solved, with the result
\be \label{L-sol}
L(t) = \left[ \gamma^2 \Delta R_S \left( \f{9 t^2}{4\gamma^2\Delta} + 1 \right) \right]^{1/3},
\ee
and $\rho = 3M / 4 \pi L^3$.

This solution shows that the star contracts until it reaches $L = x_{\rm min} = (\gamma^2 \Delta R_S)^{1/3} \sim (\lp^2 R_S)^{1/3}$ at $t=0$, when there is a bounce, and after this the radius $L$ of the star begins to expand: quantum gravity effects generate a non-singular transition from a black hole to something similar to a white hole, as has previously been suggested \cite{Rovelli:2014cta, Haggard:2014rza, Barcelo:2014cla}.  The difference between the white hole solution proposed in these works and the post-bounce $t>0$ solution derived here is subtle, but important.  In the white hole solution for vacuum general relativity, the whole interior of the white hole is anti-trapped.  For this solution, it is only part of the region inside $x < L(t)$ that is anti-trapped, while on the other hand the region $L(t) < x < R_S$ is trapped, like for a black hole; see Fig.~1.  $L$ will move outwards, and once $L$ reaches a value $L(t) > x_{\rm outer} \sim R_S$, the apparent horizon will go away and so at this time there is no longer a black hole---this Oppenheimer-Snyder space-time only contains a black hole for the time interval between when $L = x_{\rm outer}$ during the contracting phase and when $L = x_{\rm outer}$ after the bounce.

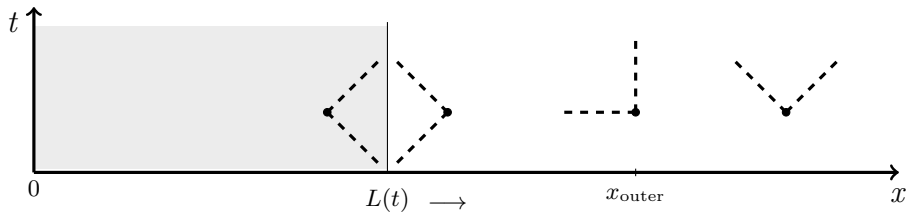
\begin{figure*}
\centering
\begin{tikzpicture}
  \fill[gray!15!white] (0.02,0.05) rectangle (4.7,1.95);
  \draw[->, very thick](0,0)--(0,2.2);
  \draw[->, very thick](0,0)--(11.5,0);
  \draw(4.7,0)--(4.7,2);
  \draw(8,-0.05)--(8,0.05);
  \draw[fill] (10,0.8) circle [radius=0.05];
  \draw[dashed, very thick](10,0.8)--(9.3,1.5);
  \draw[dashed, very thick](10,0.8)--(10.7,1.5);
  \draw[fill] (8,0.8) circle [radius=0.05];
  \draw[dashed, very thick](8,0.8)--(8,1.8);
  \draw[dashed, very thick](8,0.8)--(7,0.8);
  \draw[fill] (5.5,0.8) circle [radius=0.05];
  \draw[dashed, very thick](5.5,0.8)--(4.8,1.5);
  \draw[dashed, very thick](5.5,0.8)--(4.8,0.1);
  \draw[fill] (3.9,0.8) circle [radius=0.05];
  \draw[dashed, very thick](3.9,0.8)--(4.6,1.5);
  \draw[dashed, very thick](3.9,0.8)--(4.6,0.1);
  \node[left] at (0,2){\large{\textit{t}}$\,$};
  \node[below] at (11.5,-0.1){\large{\textit{x}}};
  \node[below] at (0,0){$0$};
  \node[below] at (4.7,-0.1){$L(t)$};
  \node[below] at (5.5,-0.25){$\longrightarrow$};
  \node[below] at (8,-0.1){$x_{\rm outer}$};
\end{tikzpicture}
\caption{This shows a snapshot of the lightcones on a spatial slice of the effective Oppenheimer-Snyder space-time at an instant of time, after the bounce has occurred and $L$ is moving outwards.  The dust field is shown in gray, and the energy density of the dust field will be larger near $L$ than in the interior.  The dust field is coupled to gravity, and will act on the space-time geometry of the outer neighbourhood of $L$ with the ultimate effect of rotating the lightcone in that neighbourhood upwards.  This in turn will allow the dust to move outwards, little by little, as the outer neighbourhood of $L$ stops being trapped.  An estimate of the rate of expansion of $L$ is given in Sec.~\ref{s.shock}.}
\end{figure*}

Importantly, the minimal value of the radius of the star is $L = x_{\rm min}$.  This fits exactly with the constraint that the vacuum solution is only valid for $x \ge x_{\rm min}$: the curvature of the space-time (in the absence of gravitational waves) must be generated by a matter field, and since $\rho$ is bounded in LQC, matter must extend out at least to the minimal radius $x_{\rm min}$ (that depends on $M$).

The metric describing this collapse and bounce of a black hole is \eqref{ltb-pg}, with $\eb = x$ and the shift vector
\be \label{shift}
N^x =
\left\{
\begin{aligned}
& - \f{6 x t}{9 t^2 + 4 \gamma^2 \Delta} \quad & \mbox{if } x \le L(t), \\
& \sqrt{ \f{R_S}{x} \left( 1 - \f{\gamma^2 \Delta R_S}{x^3} \right)} & \mbox{if } x > L(t).
\end{aligned}
\right.
\ee
For the interior, the coordinate transformation $x = L(t) \chi$ gives the flat FLRW metric
\be
ds^2=-dt^2+L^2(t)\left(d\chi^2+\chi^2d\Omega^2\right).
\ee
It is easy to verify that $N^x$ is continuous at $x=L$ when the star is contracting, but these terms differ by a sign after the bounce: for $t > 0$, there will be a shock wave in the gravitational field with a discontinuity at $x=L$ in $b(x)$ and in the effective metric.

The presence of a discontinuity in the gravitational field variable $b$ (and also in the space-time metric) shows that the assumption that edge effects can be neglected (which the calculations giving \eqref{sinb-os}--\eqref{shift} rely upon) fails after the bounce at $t=0$, and therefore the solution \eqref{sinb-os}--\eqref{shift} cannot be expected to be correct for the expanding post-bounce phase.  To correctly describe the expanding phase and the effect of the shock wave that appears then, it is necessary to include the edge effects which have been ignored so far.  Note that this shock wave, i.e., the discontinuity in $b$, will persist when edge effects are included---so the importance of including edge effects is not to remove the shock wave (which is predicted to form whether edge effects are included in the analysis or not), but rather to be able to accurately calculate the rate at which the shock wave moves outwards after the bounce.

\section{The White Hole Shock Wave}
\label{s.shock}

Since $\eb = x$, as assumed at the onset of the Oppenheimer-Snyder collapse, the dynamics of the shock wave are given by $\dot E^b = 0$, due to \eqref{eq:ebdot}, and
\begin{align} \label{bdot-edge}
\dot b &= - 4 \pi G \gamma x \rho, \\
\label{rho-edge}
\dot \rho &= \f{1}{3 \gamma \Delta x^2} \pd_x \left( x^4 N^x \rho \right),
\end{align}
which follow from \eqref{density}, \eqref{phys-ham-lqg}, \eqref{shift-gen}, and \eqref{eq:bdot}.  Note that these equations hold even if $\rho$ is not (piecewise) spatially homogeneous.

The first equation clearly shows that $b$ is monotonically decreasing, and constant outside the star where $\rho=0$ (this is why $N^x$ remains constant for $x > L$).  Away from the edge $x=L$, the second equation shows that $\rho$ increases for $N^x > 0$ (a collapsing star) while $\rho$ decreases for $N^x < 0$ (the post-bounce expanding white hole).

One way to explore the edge dynamics near $x=L$ is to do a simple discretization of \eqref{rho-edge} on a lattice with spacing $\delta x$, replacing $\pd_x f(x_i) \to [f(x_{i+1}) - f(x_{i-1})]/2\delta x$.  During collapse, $N^x > 0$ everywhere and at the edge $x_i = L$, the discretized derivative is negative since $\rho(x_{i+1}) = 0$ and $\rho(x_{i-1}) > 0$, so $\dot\rho(x_i) < 0$.  As expected, in a collapse the density increases inside the star, and decreases at $x_i$ precisely as the edge $L$ becomes smaller than $x_i$.  This agrees with \eqref{shift}, which can be trusted for the collapse phase as there is no shock wave then.

On the other hand, after the bounce $N^x < 0$ inside the expanding star, but $N^x > 0$ in the surrounding vacuum region.  If the edge is at $x_i=L$ (assuming $N^x < 0$ for $x_j \le x_i$ and $N^x > 0$ for $x_j \ge x_{i+1}$), it is easy to see that $\dot \rho(x_{i+1}) > 0$ since $\rho(x_{i+2}) = 0$ and $N^x(x_{i}) < 0$.  However, at the next time step the dust field cannot (yet) go beyond $x_{i+1}$: this is because $\dot \rho(x_{i+2}) < 0$ due to $N^x(x_{i+1}) > 0$, even though now $\rho(x_{i+1}) > 0$.  (Of course, $\rho$ cannot decrease below 0, $\dot \rho < 0$ occuring in this context is an artefact of the simple discretization.)  Instead, the white hole cannot expand further until $N^x(x_{i+1})$ becomes negative, i.e., when $\sin (\sqrt\Delta b(x_{i+1}) / x_{i+1}) = -1$.  So the discontinuity in $N^x$ will cause the white hole to expand at a much slower rate than it collapsed.  Also, the dust field will accumulate near the edge $x=L$ (since $\dot \rho(x_i),~ \dot\rho(x_{i+1}) > 0$ as long as $L$ does not move), so $\rho$ will become greater at the wave front than inside the star.  Results obtained neglecting edge effects can be trusted far from the white hole edge, but the wave front location $L$ will move outwards at a slowed rate and $\rho$ will be greater near $L$ than for $x \ll L$.

From a physical perspective, the dust field in the vicinity of $x=L$ is caught between an outside trapped region and an inside anti-trapped region, as shown in Fig.~1.  Although the matter field located at $L$ initially cannot move, it is of course coupled to gravity and changes the geometry of the space-time in its neighbourhood.  In particular, in the outer neighbourhood of $L$, the orientation of the lightcones will change so that this region is no longer trapped, and then the dust field will be able to move outwards into this region.  This slow process will end once $L$ reaches the outer horizon, at which point the matter fields will be able to freely move outwards.  In the following, we estimate the time required for $L$ to reach $x_{\rm outer}$.

This calculation can be used to obtain an estimate for the lifetime $T$ of a black hole (as measured by a distant observer detecting light signals emitted from the surface of the star), which is given by the coordinate time $t$ elapsed between the instants when $L = x_{\rm outer} \sim R_S$ before and after the bounce, as shown in the Appendix.

From \eqref{L-sol}, the duration of the contracting portion (from $L = x_{\rm outer}$ to $L=x_{\rm min}$) is $\sim R_S$, so the last step to find $T$ is to calculate the duration of the expanding phase, from $L = x_{\rm min}$ to when $L = x_{\rm outer}$ once more.  A precise determination of the white hole's lifetime $T$ will require a considerably more detailed analysis, likely including high-resolution numerics, but it is possible to obtain a simple estimate for $T$ by using \eqref{bdot-edge} to calculate the time $\delta t_i$, for each location of the shock-wave front $L$, it takes for $N^x$ to change sign, and then sum over all $x$ from $x_{\rm min}$ to $R_S$, assuming $\delta x \sim \lp$.

It is enough to start the calculation at, say, $x_i \sim \sqrt{\lp R_S} \gg x_{\rm min}$.  (The following calculation can also be used to estimate the contribution to $T$ for the time elapsed while $x_{\rm min} < L(t) < x_i$, which shows that it is subleading compared to the contribution from $x_i < L < x_{\rm outer}$ and so can safely be ignored for an estimate of the leading contribution to $T$.) When the front $L$ of the shock wave first reaches the radius $x$, then $|\sin (\sqrt\Delta b / x)| \ll 1$ for $x \ge x_i$ and therefore, for $N^x$ to change signs, $b$ must change by $\sim -\pi x / 2 \sqrt\Delta$.  Then, from \eqref{bdot-edge} it follows that (dropping numerical prefactors of order 1) $\delta t_i \sim 1/(G \sqrt\Delta \rho(x_i))$.

At the bounce, $\rho = \rho_c$ but $\rho$ will decrease as the shock wave expands.  The leading edge of the shock wave will have a greater $\rho$ than the center (where $\rho$ evolves in a symmetric fashion around the bounce) since the dust field will be pushed towards the edge where it will accumulate due to the slow expansion of the front of the shock wave.  After some time, it can be expected that a significant fraction of the dust field will lie within a short distance $w$ of the leading front, in which case the energy density at the edge $\rho_e$ will scale as $\rho_e \sim M / (x^2 w)$.  Evaluating $\rho$ at the bounce gives $\rho_c \sim M / x_{\rm min}^3$ so $M \sim \Delta \rho_c R_S$ and $\rho_e \sim \Delta \rho_c R_s / x^2 w$.  Then, $\delta t_i \sim w x^2/(G \Delta^{3/2} R_S \rho_c)$ and, since the pre-bounce phase is much shorter than the post-bounce expansion,
\be
T \sim \sum_i \delta t_i \sim \int_{x_i}^{R_S} \f{\delta t_i}{\lp} dx \sim \f{w R_S^2}{\lp^2},
\ee
similar what is suggested in \cite{Christodoulou:2016vny}. If $w$ is independent of $R_S$, then $w/\lp$ is a (potentially large) dimensionless constant and the black hole lifetime $T \sim R_S^2/\lp$ is significantly shorter than the Page time \cite{Page:1993df} in which case the standard black hole information loss problem is avoided.

\section{Discussion}

The LTB space-times are spherically symmetric space-times coupled to pressureless dust.  After imposing some convenient gauges, we constructed an effective Hamiltonian following the standard LQC procedure of replacing components of the Ashtekar-Barbero connection by holonomies using the $\bar\mu$ scheme, and derived the effective equations for LTB space-times.  A particularly interesting case is the Oppenheimer-Snyder model for black hole collapse, which can be solved exactly if edge effects are neglected; the result is that the star contracts until the density reaches the LQC critical density $\rho_c \sim \rho_{\rm Pl}$ and then bounces; for the post-bounce phase it is necessary to include edge effects which become important.  Although the post-bounce phase does not exactly correspond to a white hole, there are some significant similarities which lead us to view this model as a specific realization of quantum gravity generating a non-singular transition from a black hole to a `white hole', as proposed in \cite{Rovelli:2014cta, Haggard:2014rza, Barcelo:2014cla}, with this general picture also found in \cite{Bambi:2013caa, Liu:2014kra, BenAchour:2020gon, Schmitz:2019jct, Piechocki:2020bfo}.  Also, since the effective equations in this case are known for both the interior and exterior regions, as well as at the discontinuity at the surface of the star, it is now possible to properly include the edge effects that become large after the bounce.

The dynamics of the LQG Oppenheimer-Snyder space-time can be split in two main phases: the black hole collapse and the white hole shock wave.  First, the star collapses until its surface $L(t)$ reaches $x_{\rm outer} \sim R_S$, at which point an apparent horizon appears and the star forms a black hole.  Then, quantum gravity effects become important when $L$ nears $x_{\rm min} \sim (\lp^2 R_S)^{1/3}$.  The quantum gravity effects are sufficiently strong to stop the collapse of the star and cause a bounce.  Note that the region inside $x_{\rm min}$ is never trapped as there is an interior horizon at $x = x_{\rm inner} \sim x_{\rm min} + (\gamma^4 \Delta^2/27 R_S)^{1/3}$.  Then, after the bounce, edge effects become important and slow the expansion of the star.  These edge effects can be understood qualitatively to arise from the surface $L$ being caught between an anti-trapped region lying within part of the interior of the star, and the trapped region outside the star for $L < x < x_{\rm outer}$.  As the surface of the star $L$ slowly moves outwards, the trapped region becomes smaller, until $L = x_{\rm outer}$ and the apparent horizon vanishes; at this point the solution no longer corresponds to a black hole.  We estimate that due to these edge effects, the lifetime of the black hole (as measured by a distant observer detecting light signals emitted from the surface of the star) is $T \sim R_S^2/\lp$. Note that matter plays an important role in this context, giving a much shorter lifetime than what spin foam calculations have found for vacuum space-times \cite{Christodoulou:2018ryl, Bianchi:2018}.  After the surface $L$ of the star has once more passed $x = x_{\rm outer}$, high-energy photons could be emitted by the expanding star, producing a potentially observable astronomical signature.

There are two results that show a remarkable unity between LQG holonomy corrections applied to different families of space-times.  First, the equation of motion for the Oppenheimer-Snyder (flat FLRW) interior is identical to the LQC effective Friedmann equation for flat FLRW space-times in the limit that edge effects are neglected.  Although the derivation of the two equations is quite different, they describe the same physics, so it is reassuring that both procedures give the same dynamics.  Second, the vacuum spherically symmetric solution for a black hole of mass $M$ is only valid to a minimal radius $x_{\rm min} \sim (R_S \lp^2)^{1/3}$ \cite{Gambini:2020nsf, Kelly:2020uwj}.  This vacuum solution corresponds to the exterior of the Oppenheimer-Snyder solution, and it turns out that the minimal radius of the interior is exactly $x_{\rm min}$: to generate a Schwarzschild-like exterior of mass $M$, there must be a matter field extending to at least the radius $x_{\rm min}$.  Once again, the consistency between LQG results in vacuum and LTB space-times is remarkable.

There remain a number of important open problems that we leave for future work.  First, a better understanding of the physics of the white hole shock wave could be used to calculate the lifetime of a black hole more accurately, and could also give predictions concerning the light emitted by a white hole that could be seen by distant observers.  It would also be interesting to study more realistic black hole collapse models by including matter fields with pressure and/or a radially-varying density, and also to determine the stability of the solution to further infalling matter.  We point out that since the expanding solution that we obtain is not exactly a white hole (although it does share some qualitative similarities), the instability results for white holes \cite{Eardley:1974zz, Barcelo:2015uff} are not directly applicable here.  Nonetheless, despite their differences, it does seem likely that at least some of the phenomenology of this model could be similar to what has been found for transitions from a black hole to a white hole \cite{Barrau:2015uca, Vidotto:2018wvr, Carballo-Rubio:2018jzw}.  Finally, it will also be important to include other quantum effects, most notably Hawking radiation, to obtain a complete picture of quantum gravity effects in black hole space-times.

\appendix

\section{Lifetime of a Quantum Black Hole}

The lifetime of the black hole is of course observer-dependent.  Here we consider the lifetime as measured by a distant observer (located at a fixed radius $x = R \gg R_S$), who measures the proper time elapsed between receiving two light signals; the first sent during the collapse of the star just before the black hole forms, and the second sent shortly after the surface of the star $L$ exits the apparent horizon after the bounce.

Here, for the sake of concreteness we assume that the two light signals are emitted from the $x=L$ surface of the Oppenheimer-Snyder star when $L=3GM$, but other choices will give similar results.  Explicitly, the first light signal is emitted when $L=3GM$ during the contracting phase, and the second light signal is emitted when $L=3GM$ after the bounce.

Using the coordinate system \eqref{metric} for the vacuum exterior of the star, null radial geodesics satisfy
\be
0 = -\dot t^2 + 2 N^x \dot t \dot x + \dot x^2,
\ee
with dots denoting a derivative with respect to an affine parameter.
Note that the explicit form of $N^x$ is not necessary here, all that matters is that $N^x$ depends only the radial coordinate $x$ but not on the time coordinate $t$; this condition will always be satisfied for the exterior of the Oppenheimer-Snyder star in spherical symmetry, even without assuming the Einstein equations (in classical general relativity, $N^x = \sqrt{R_S/x}$).

By solving for $\dot t / \dot x = dt/dx$ and integrating, it follows that the $t$ and $x$ coordinates for outgoing null radial geodesics (passing through $x=x_o > R_S$ at the coordinate time $t=t_o$) satisfy
\be \label{geod}
t(x) = t_o + f(x) - f(x_o),
\ee
with $f(x) = \int \! dx \, [N^x + \sqrt{(N^x)^2 + 1 \,} \, ]$.

Therefore, if two light signals are emitted at $x=3GM$ at times $t_1(3GM)=0$ and $t_2(3GM) = \Delta t_L$, then the elapsed time (as measured by a distant observer located at $x = R \gg R_S$) will be, using \eqref{geod},
\begin{align}
\Delta t_{\rm d} &= t_2(R) - t_1(R) \nn \\
&=t_2(3GM) + f(R) - f(3GM) \nn \\
& \qquad - [t_1(3GM) + f(R) - f(3GM)] \nn \\
& = \Delta t_L.
\end{align}
Finally, the elapsed proper time for an observer at constant radius $x=R \gg R_S$ is $\Delta\tau_{\rm d} \approx \Delta t_{\rm d}$ for the metric \eqref{metric} (up to small corrections of the order $R_S/R$ that can safely be neglected), therefore
\be
\Delta \tau_{\rm d} \approx \Delta t_{\rm d} = \Delta t_L.
\ee
So the lifetime of a quantum black hole is simply given by the difference in coordinate time, in terms of the metric \eqref{metric}, between the times the first and second light signals were emitted for some $x_{\rm emission} > R_S$, respectively from the collapse and from the shock wave after the bounce.

\acknowledgments
We thank Viqar Husain for helpful discussions.
This work was supported in part by the Natural Sciences and Engineering Research Council of Canada.

\raggedright


\begin{thebibliography}{10}

\bibitem{Lemaitre:1933gd}
G.~Lema{\^i}tre, ``{The expanding universe},'' Ann. Soc. Sci. Bruxelles {\bf A53}
  (1933) 51. Republished in Gen. Rel. Grav. {\bf 29} (1997) 641.

\bibitem{Tolman:1934za}
R.~C. Tolman, ``{Effect of inhomogeneity on cosmological models},'' Proc. Nat.
  Acad. Sci. {\bf 20} (1934) 169--176. Republished in Gen. Rel. Grav. {\bf 29}
  (1997) 935.

\bibitem{Bondi:1947fta}
H.~Bondi, ``{Spherically symmetrical models in general relativity},'' Mon. Not.
  Roy. Astron. Soc. {\bf 107} (1947) 410--425.

\bibitem{Thiemann:2007pyv}
T.~Thiemann, {\em {Modern Canonical Quantum General Relativity}}.
\newblock Cambridge Monographs on Mathematical Physics. Cambridge University
  Press, 2007.

\bibitem{Vaz:2000zb}
C.~Vaz, L.~Witten and T.~Singh, ``{Toward a midisuperspace quantization of
  Lemaitre-Tolman-Bondi collapse models},'' Phys. Rev. D \textbf{63} (2001),
  104020,
\href{http://arXiv.org/abs/gr-qc/0012053}{{\tt arXiv:gr-qc/0012053}}.

\bibitem{Kiefer:2005tw}
C.~Kiefer, J.~Muller-Hill and C.~Vaz, ``{Classical and quantum LTB model for
  the non-marginal case},'' Phys. Rev. D \textbf{73} (2006) 044025,
\href{http://arXiv.org/abs/gr-qc/0512047}{{\tt arXiv:gr-qc/0512047}}.

\bibitem{Kiefer:2019csi}
C.~Kiefer and T.~Schmitz, ``{Singularity avoidance for collapsing quantum dust in
  the Lema{\^i}tre-Tolman-Bondi model},'' Phys. Rev. D \textbf{99} (2019) 126010,
\href{http://arXiv.org/abs/1904.13220}{{\tt arXiv:1904.13220}}.

\bibitem{Ashtekar:2011ni}
A.~Ashtekar and P.~Singh, ``{Loop Quantum Cosmology: A Status Report},'' Class.
  Quant. Grav. {\bf 28} (2011) 213001,
\href{http://arXiv.org/abs/1108.0893}{{\tt arXiv:1108.0893}}.
%%CITATION = ARXIV:1108.0893;%%.

\bibitem{Ashtekar:2006wn}
A.~Ashtekar, T.~Pawlowski, and P.~Singh, ``{Quantum Nature of the Big Bang:
  Improved dynamics},'' Phys. Rev. {\bf D74} (2006) 084003,
\href{http://arXiv.org/abs/gr-qc/0607039}{{\tt arXiv:gr-qc/0607039}}.
%%CITATION = GR-QC/0607039;%%.

\bibitem{Modesto:2004xx}
L.~Modesto, ``{Disappearance of black hole singularity in quantum gravity},''
  Phys. Rev. {\bf D70} (2004) 124009,
\href{http://arXiv.org/abs/gr-qc/0407097}{{\tt arXiv:gr-qc/0407097}}.
%%CITATION = GR-QC/0407097;%%.

\bibitem{Ashtekar:2005qt}
A.~Ashtekar and M.~Bojowald, ``{Quantum geometry and the Schwarzschild
  singularity},'' Class. Quant. Grav. {\bf 23} (2006) 391--411,
\href{http://arXiv.org/abs/gr-qc/0509075}{{\tt arXiv:gr-qc/0509075}}.
%%CITATION = GR-QC/0509075;%%.

\bibitem{Bohmer:2007wi}
C.~G. Boehmer and K.~Vandersloot, ``{Loop Quantum Dynamics of the Schwarzschild
  Interior},'' Phys. Rev. {\bf D76} (2007) 104030,
\href{http://arXiv.org/abs/0709.2129}{{\tt arXiv:0709.2129}}.
%%CITATION = ARXIV:0709.2129;%%.

\bibitem{Campiglia:2007pb}
M.~Campiglia, R.~Gambini, and J.~Pullin, ``{Loop quantization of spherically
  symmetric midi-superspaces: The Interior problem},'' AIP Conf. Proc. {\bf
  977} (2008) 52--63,
\href{http://arXiv.org/abs/0712.0817}{{\tt arXiv:0712.0817}}.
%%CITATION = ARXIV:0712.0817;%%.

\bibitem{Chiou:2008nm}
D.-W. Chiou, ``{Phenomenological loop quantum geometry of the Schwarzschild
  black hole},'' Phys. Rev. {\bf D78} (2008) 064040,
\href{http://arXiv.org/abs/0807.0665}{{\tt arXiv:0807.0665}}.
%%CITATION = ARXIV:0807.0665;%%.

\bibitem{Brannlund:2008iw}
J.~Brannlund, S.~Kloster, and A.~DeBenedictis, ``{The Evolution of Lambda Black
  Holes in the Mini-Superspace Approximation of Loop Quantum Gravity},'' Phys.
  Rev. D {\bf 79} (2009) 084023, \href{http://arXiv.org/abs/0901.0010}{{\tt
  arXiv:0901.0010}}.

\bibitem{Joe:2014tca}
A.~Joe and P.~Singh, ``{Kantowski-Sachs spacetime in loop quantum cosmology:
  bounds on expansion and shear scalars and the viability of quantization
  prescriptions},'' Class. Quant. Grav. {\bf 32} (2015) 015009,
  \href{http://arXiv.org/abs/1407.2428}{{\tt arXiv:1407.2428}}.

\bibitem{Corichi:2015xia}
A.~Corichi and P.~Singh, ``{Loop quantization of the Schwarzschild interior
  revisited},'' Class. Quant. Grav. {\bf 33} (2016) 055006,
\href{http://arXiv.org/abs/1506.08015}{{\tt arXiv:1506.08015}}.
%%CITATION = ARXIV:1506.08015;%%.

\bibitem{Cortez:2017alh}
J.~Cortez, W.~Cuervo, H.~A. Morales-T{\'e}cotl, and J.~C. Ruelas, ``{Effective
  loop quantum geometry of Schwarzschild interior},'' Phys. Rev. D {\bf 95}
  (2017) 064041, \href{http://arXiv.org/abs/1704.03362}{{\tt
  arXiv:1704.03362}}.

\bibitem{Olmedo:2017lvt}
J.~Olmedo, S.~Saini, and P.~Singh, ``{From black holes to white holes: a
  quantum gravitational, symmetric bounce},'' Class. Quant. Grav. {\bf 34}
  (2017) 225011, \href{http://arXiv.org/abs/1707.07333}{{\tt
  arXiv:1707.07333}}.

\bibitem{BenAchour:2018khr}
J.~Ben~Achour, F.~Lamy, H.~Liu, and K.~Noui, ``{Polymer Schwarzschild black
  hole: An effective metric},'' EPL {\bf 123} (2018) 20006,
\href{http://arXiv.org/abs/1803.01152}{{\tt arXiv:1803.01152}}.
%%CITATION = ARXIV:1803.01152;%%.

\bibitem{Ashtekar:2018lag}
A.~Ashtekar, J.~Olmedo, and P.~Singh, ``{Quantum Transfiguration of Kruskal
  Black Holes},'' Phys. Rev. Lett. {\bf 121} (2018) 241301,
\href{http://arXiv.org/abs/1806.00648}{{\tt arXiv:1806.00648}}.
%%CITATION = ARXIV:1806.00648;%%.

\bibitem{Bodendorfer:2019cyv}
N.~Bodendorfer, F.~M. Mele, and J.~M{\"u}nch, ``{Effective Quantum Extended
  Spacetime of Polymer Schwarzschild Black Hole},'' Class. Quant. Grav. {\bf
  36} (2019) 195015,
\href{http://arXiv.org/abs/1902.04542}{{\tt arXiv:1902.04542}}.
%%CITATION = ARXIV:1902.04542;%%.

\bibitem{Alesci:2019pbs}
E.~Alesci, S.~Bahrami, and D.~Pranzetti, ``{Quantum gravity predictions for
  black hole interior geometry},'' Phys. Lett. {\bf B797} (2019) 134908,
\href{http://arXiv.org/abs/1904.12412}{{\tt arXiv:1904.12412}}.
%%CITATION = ARXIV:1904.12412;%%.

\bibitem{Assanioussi:2019twp}
M.~Assanioussi, A.~Dapor, and K.~Liegener, ``{Perspectives on the dynamics in a
  loop quantum gravity effective description of black hole interiors},'' Phys.
  Rev. D {\bf 101} (2020) 026002,
  \href{http://arXiv.org/abs/1908.05756}{{\tt arXiv:1908.05756}}.

\bibitem{Bojowald:2005cb}
M.~Bojowald and R.~Swiderski, ``{Spherically symmetric quantum geometry:
  Hamiltonian constraint},'' Class. Quant. Grav. {\bf 23} (2006) 2129--2154,
\href{http://arXiv.org/abs/gr-qc/0511108}{{\tt arXiv:gr-qc/0511108}}.
%%CITATION = GR-QC/0511108;%%.

\bibitem{Gambini:2008dy}
R.~Gambini and J.~Pullin, ``{Black holes in loop quantum gravity: The Complete
  space-time},'' Phys. Rev. Lett. {\bf 101} (2008) 161301,
\href{http://arXiv.org/abs/0805.1187}{{\tt arXiv:0805.1187}}.
%%CITATION = ARXIV:0805.1187;%%.

\bibitem{Reyes:2009}
J.~D. Reyes, ``{Spherically Symmetric Loop Quantum Gravity: Connection to
  Two-Dimensional Models and Applications to Gravitational Collapse},''. PhD
  thesis, The Pennsylvania State University, 2009. Available online at
  \href{https://etda.libraries.psu.edu/catalog/10349}{https://etda.libraries.psu.edu/catalog/10349}.

\bibitem{Gambini:2013ooa}
R.~Gambini and J.~Pullin, ``{Loop quantization of the Schwarzschild black
  hole},'' Phys. Rev. Lett. {\bf 110} (2013) 211301,
\href{http://arXiv.org/abs/1302.5265}{{\tt arXiv:1302.5265}}.
%%CITATION = ARXIV:1302.5265;%%.

\bibitem{Gambini:2013hna}
R.~Gambini, J.~Olmedo, and J.~Pullin, ``{Quantum black holes in Loop Quantum
  Gravity},'' Class. Quant. Grav. {\bf 31} (2014) 095009,
\href{http://arXiv.org/abs/1310.5996}{{\tt arXiv:1310.5996}}.
%%CITATION = ARXIV:1310.5996;%%.

\bibitem{BenAchour:2016brs}
J.~Ben~Achour, S.~Brahma, and A.~Marciano, ``{Spherically symmetric sector of
  self dual Ashtekar gravity coupled to matter: Anomaly-free algebra of
  constraints with holonomy corrections},'' Phys. Rev. {\bf D96} (2017)
  026002,
\href{http://arXiv.org/abs/1608.07314}{{\tt arXiv:1608.07314}}.
%%CITATION = ARXIV:1608.07314;%%.

\bibitem{Bojowald:2018xxu}
M.~Bojowald, S.~Brahma, and D.-h. Yeom, ``{Effective line elements and
  black-hole models in canonical loop quantum gravity},'' Phys. Rev. {\bf D98}
  (2018) 046015,
\href{http://arXiv.org/abs/1803.01119}{{\tt arXiv:1803.01119}}.
%%CITATION = ARXIV:1803.01119;%%.

\bibitem{Chiou:2012pg}
D.-W. Chiou, W.-T. Ni, and A.~Tang, ``{Loop quantization of spherically
  symmetric midisuperspaces and loop quantum geometry of the maximally extended
  Schwarzschild spacetime},''
\href{http://arXiv.org/abs/1212.1265}{{\tt arXiv:1212.1265}}.
%%CITATION = ARXIV:1212.1265;%%.

\bibitem{Gambini:2020nsf}
R.~Gambini, J.~Olmedo, and J.~Pullin, ``{Spherically symmetric loop quantum
  gravity: analysis of improved dynamics},''
\href{http://arXiv.org/abs/2006.01513}{{\tt arXiv:2006.01513}}.

\bibitem{Kelly:2020uwj}
J.~G.~Kelly, R.~Santacruz and E.~Wilson-Ewing, ``{Effective loop quantum
  gravity framework for vacuum spherically symmetric space-times},''
\href{http://arXiv.org/abs/2006.09302}{{\tt arXiv:2006.09302}}.

\bibitem{Gambini:2009ie}
R.~Gambini, J.~Pullin, and S.~Rastgoo, ``{Quantum scalar field in quantum
  gravity: The vacuum in the spherically symmetric case},'' Class. Quant. Grav.
  {\bf 26} (2009) 215011,
\href{http://arXiv.org/abs/0906.1774}{{\tt arXiv:0906.1774}}.
%%CITATION = ARXIV:0906.1774;%%.

\bibitem{Gambini:2014qta}
R.~Gambini, E.~M. Capurro, and J.~Pullin, ``{Quantum spacetime of a charged
  black hole},'' Phys. Rev. {\bf D91} (2015) 084006,
\href{http://arXiv.org/abs/1412.6055}{{\tt arXiv:1412.6055}}.
%%CITATION = ARXIV:1412.6055;%%.

\bibitem{Bojowald:2015zha}
M.~Bojowald, S.~Brahma, and J.~D. Reyes, ``{Covariance in models of loop
  quantum gravity: Spherical symmetry},'' Phys. Rev. {\bf D92} (2015)
  045043,
\href{http://arXiv.org/abs/1507.00329}{{\tt arXiv:1507.00329}}.
%%CITATION = ARXIV:1507.00329;%%.

\bibitem{Campiglia:2016fzp}
M.~Campiglia, R.~Gambini, J.~Olmedo, and J.~Pullin, ``{Quantum self-gravitating
  collapsing matter in a quantum geometry},'' Class. Quant. Grav. {\bf 33}
  (2016) 18LT01,
\href{http://arXiv.org/abs/1601.05688}{{\tt arXiv:1601.05688}}.
%%CITATION = ARXIV:1601.05688;%%.

\bibitem{Husain:2006cx}
V.~Husain and O.~Winkler, ``{Quantum Hamiltonian for gravitational collapse},''
  Phys. Rev. D {\bf 73} (2006) 124007,
  \href{http://arXiv.org/abs/gr-qc/0601082}{{\tt arXiv:gr-qc/0601082}}.

\bibitem{Husain:2008tc}
V.~Husain, ``{Critical behaviour in quantum gravitational collapse},'' Adv.
  Sci. Lett. {\bf 2} (2009) 214,
\href{http://arXiv.org/abs/0808.0949}{{\tt arXiv:0808.0949}}.
%%CITATION = ARXIV:0808.0949;%%.

\bibitem{Hossenfelder:2009fc}
S.~Hossenfelder, L.~Modesto, and I.~Premont-Schwarz, ``{A Model for
  non-singular black hole collapse and evaporation},'' Phys. Rev. {\bf D81}
  (2010) 044036,
\href{http://arXiv.org/abs/0912.1823}{{\tt arXiv:0912.1823}}.
%%CITATION = ARXIV:0912.1823;%%.

\bibitem{Benitez:2020szx}
F.~Benitez, R.~Gambini, L.~Lehner, S.~Liebling, and J.~Pullin, ``{Critical
  collapse of a scalar field in semiclassical loop quantum gravity},'' Phys.
  Rev. Lett. {\bf 124} (2020) 071301,
\href{http://arXiv.org/abs/2002.04044}{{\tt arXiv:2002.04044}}.
%%CITATION = ARXIV:2002.04044;%%.

\bibitem{Ashtekar:2008jd}
A.~Ashtekar, V.~Taveras, and M.~Varadarajan, ``{Information is Not Lost in the
  Evaporation of 2-dimensional Black Holes},'' Phys. Rev. Lett. {\bf 100}
  (2008) 211302, \href{http://arXiv.org/abs/0801.1811}{{\tt arXiv:0801.1811}}.

\bibitem{Ashtekar:2010qz}
A.~Ashtekar, F.~Pretorius, and F.~M. Ramazanoglu, ``{Evaporation of
  2-Dimensional Black Holes},'' Phys. Rev. D {\bf 83} (2011) 044040,
  \href{http://arXiv.org/abs/1012.0077}{{\tt arXiv:1012.0077}}.

\bibitem{Tavakoli:2013rna}
Y.~Tavakoli, J.~Marto, and A.~Dapor, ``{Semiclassical dynamics of horizons in
  spherically symmetric collapse},'' Int. J. Mod. Phys. D {\bf 23} (2014)
  1450061, \href{http://arXiv.org/abs/1303.6157}{{\tt arXiv:1303.6157}}.

\bibitem{Bambi:2013caa}
C.~Bambi, D.~Malafarina and L.~Modesto, ``{Non-singular quantum-inspired
  gravitational collapse},'' Phys. Rev. D \textbf{88} (2013), 044009,
  \href{http://arXiv.org/abs/1305.4790}{{\tt arXiv:1305.4790}}.

\bibitem{Liu:2014kra}
Y.~Liu, D.~Malafarina, L.~Modesto and C.~Bambi, ``{Singularity avoidance in
  quantum-inspired inhomogeneous dust collapse},'' Phys. Rev. D \textbf{90}
  (2014) 044040,
  \href{http://arXiv.org/abs/1405.7249}{{\tt arXiv:1405.7249}}.

\bibitem{Christodoulou:2016vny}
M.~Christodoulou, C.~Rovelli, S.~Speziale, and I.~Vilensky, ``{Planck star
  tunneling time: An astrophysically relevant observable from background-free
  quantum gravity},'' Phys. Rev. D {\bf 94} (2016) 084035,
  \href{http://arXiv.org/abs/1605.05268}{{\tt arXiv:1605.05268}}.

\bibitem{Christodoulou:2018ryl}
M.~Christodoulou and F.~D'Ambrosio, ``{Characteristic Time Scales for the
  Geometry Transition of a Black Hole to a White Hole from Spinfoams},''
  \href{http://arXiv.org/abs/1801.03027}{{\tt arXiv:1801.03027}}.

\bibitem{BenAchour:2020gon}
J.~Ben~Achour, S.~Brahma, S.~Mukohyama, and J.-P. Uzan, ``{Consistent
  black-to-white hole bounces from matter collapse},''
  \href{http://arXiv.org/abs/2004.12977}{{\tt arXiv:2004.12977}}.

\bibitem{Bianchi:2018}
E.~Bianchi, M.~Christodoulou, F.~D'Ambrosio, H.~Haggard, C.~Rovelli, ``{White
  Holes as Remnants: A Surprising Scenario for the End of a Black Hole},''
  Class. Quant. Grav. \textbf{35} (2018) 225003,
\href{https://arxiv.org/abs/1802.04264}{{\tt arXiv:1802.04264}}.

\bibitem{Husain:2004yz}
V.~Husain and O.~Winkler, ``{Quantum resolution of black hole singularities},''
  Class. Quant. Grav. {\bf 22} (2005) L127--L134,
\href{http://arXiv.org/abs/gr-qc/0410125}{{\tt arXiv:gr-qc/0410125}}.
%%CITATION = GR-QC/0410125;%%.

\bibitem{Ziprick:2009nd}
J.~Ziprick and G.~Kunstatter, ``{Dynamical Singularity Resolution in
  Spherically Symmetric Black Hole Formation},'' Phys. Rev. D {\bf 80} (2009)
  024032, \href{http://arXiv.org/abs/0902.3224}{{\tt arXiv:0902.3224}}.

\bibitem{Bojowald:2009ih}
M.~Bojowald, J.~D. Reyes, and R.~Tibrewala, ``{Non-marginal LTB-like models
  with inverse triad corrections from loop quantum gravity},'' Phys. Rev. D
  {\bf 80} (2009) 084002, \href{http://arXiv.org/abs/0906.4767}{{\tt
  arXiv:0906.4767}}.

\bibitem{Kreienbuehl:2010vc}
A.~Kreienbuehl, V.~Husain, and S.~S. Seahra, ``{Modified general relativity as
  a model for quantum gravitational collapse},'' Class. Quant. Grav. {\bf 29}
  (2012) 095008, \href{http://arXiv.org/abs/1011.2381}{{\tt arXiv:1011.2381}}.

\bibitem{Bojowald:2011js}
M.~Bojowald, G.~M. Paily, J.~D. Reyes, and R.~Tibrewala, ``{Black-hole horizons
  in modified space-time structures arising from canonical quantum gravity},''
  Class. Quant. Grav. {\bf 28} (2011) 185006,
\href{http://arXiv.org/abs/1105.1340}{{\tt arXiv:1105.1340}}.
%%CITATION = ARXIV:1105.1340;%%.

\bibitem{Husain:2011tk}
V.~Husain and T.~Pawlowski, ``{Time and a physical Hamiltonian for quantum
  gravity},'' Phys. Rev. Lett. {\bf 108} (2012) 141301,
\href{http://arXiv.org/abs/1108.1145}{{\tt arXiv:1108.1145}}.
%%CITATION = ARXIV:1108.1145;%%.

\bibitem{Aruga:2019dwq}
D.~Arruga, J.~Ben Achour and K.~Noui, ``{Deformed General Relativity and Quantum
  Black Holes Interior},'' Universe \textbf{6} (2020) 39,
\href{https://arxiv.org/abs/1912.02459}{{\tt arXiv:1912.02459}}.

\bibitem{Lasky:2006hq}
P.~D. Lasky, A.~W. Lun, and R.~B. Burston, ``{Initial value formalism for dust
  collapse},'' {ANZIAM Journal} {\bf 49} (2007) 205,
  \href{http://arXiv.org/abs/gr-qc/0606003}{{\tt arXiv:gr-qc/0606003}}.

\bibitem{Giesel:2009jp}
K.~Giesel, J.~Tambornino, and T.~Thiemann, ``{LTB spacetimes in terms of Dirac
  observables},'' Class. Quant. Grav. {\bf 27} (2010) 105013,
\href{http://arXiv.org/abs/0906.0569}{{\tt arXiv:0906.0569}}.
%%CITATION = ARXIV:0906.0569;%%.

\bibitem{Taveras:2008ke}
V.~Taveras, ``{Corrections to the Friedmann Equations from LQG for a Universe
  with a Free Scalar Field},'' Phys. Rev. D {\bf 78} (2008) 064072,
  \href{http://arXiv.org/abs/0807.3325}{{\tt arXiv:0807.3325}}.

\bibitem{Rovelli:2013zaa}
C.~Rovelli and E.~Wilson-Ewing, ``{Why are the effective equations of loop
  quantum cosmology so accurate?},'' Phys. Rev. D {\bf 90} (2014)
  023538, \href{http://arXiv.org/abs/1310.8654}{{\tt arXiv:1310.8654}}.

\bibitem{Vandersloot:2006ws}
K.~Vandersloot, ``{Loop quantum cosmology and the k = - 1 RW model},'' Phys.
  Rev. D {\bf 75} (2007) 023523, \href{http://arXiv.org/abs/gr-qc/0612070}{{\tt
  arXiv:gr-qc/0612070}}.

\bibitem{Singh:2013ava}
P.~Singh and E.~Wilson-Ewing, ``{Quantization ambiguities and bounds on
  geometric scalars in anisotropic loop quantum cosmology},'' Class. Quant.
  Grav. {\bf 31} (2014) 035010,
\href{http://arXiv.org/abs/1310.6728}{{\tt arXiv:1310.6728}}.
%%CITATION = ARXIV:1310.6728;%%.

\bibitem{Ashtekar:2009um}
A.~Ashtekar and E.~Wilson-Ewing, ``{Loop quantum cosmology of Bianchi type II
  models},'' Phys. Rev. D {\bf 80} (2009) 123532,
  \href{http://arXiv.org/abs/0910.1278}{{\tt arXiv:0910.1278}}.

\bibitem{Oppenheimer:1939ue}
J.~R. Oppenheimer and H.~Snyder, ``{On Continued gravitational contraction},''
  Phys. Rev. {\bf 56} (1939)
455--459.
%%CITATION = PHRVA,56,455;%%.

\bibitem{Rovelli:2014cta}
C.~Rovelli and F.~Vidotto, ``{Planck stars},'' Int. J. Mod. Phys. {\bf D23}
  (2014) 1442026,
\href{http://arXiv.org/abs/1401.6562}{{\tt arXiv:1401.6562}}.
%%CITATION = ARXIV:1401.6562;%%.

\bibitem{Haggard:2014rza}
H.~M. Haggard and C.~Rovelli, ``{Quantum-gravity effects outside the horizon
  spark black to white hole tunneling},'' Phys. Rev. {\bf D92} (2015)
  104020,
\href{http://arXiv.org/abs/1407.0989}{{\tt arXiv:1407.0989}}.
%%CITATION = ARXIV:1407.0989;%%.

\bibitem{Barcelo:2014cla}
C.~Barcelo, R.~Carballo-Rubio, L.~J.~Garay and G.~Jannes, ``{The lifetime problem
  of evaporating black holes: mutiny or resignation},'' Class. Quant. Grav. \textbf{32}
  (2015) 035012,
\href{http://arXiv.org/abs/1409.1501}{{\tt arXiv:1409.1501}}.
%%CITATION = ARXIV:1409.1501;%%.

\bibitem{Page:1993df}
D.~N. Page, ``{Average entropy of a subsystem},'' Phys. Rev. Lett. {\bf 71}
  (1993) 1291--1294,
\href{http://arXiv.org/abs/gr-qc/9305007}{{\tt arXiv:gr-qc/9305007}}.

\bibitem{Schmitz:2019jct}
T.~Schmitz, ``{Towards a quantum Oppenheimer-Snyder model},'' Phys. Rev. D
  \textbf{101} (2020) 026016,
\href{http://arXiv.org/abs/1912.08175}{{\tt arXiv:1912.08175}}.

\bibitem{Piechocki:2020bfo}
W.~Piechocki and T.~Schmitz, ``{Quantum Oppenheimer-Snyder model},''
\href{http://arXiv.org/abs/2004.02939}{{\tt arXiv:2004.02939}}.

\bibitem{Eardley:1974zz}
D.~M. Eardley, ``{Death of White Holes in the Early Universe},'' Phys. Rev.
  Lett. {\bf 33} (1974) 442--444.

\bibitem{Barcelo:2015uff}
C.~Barcel{\'o}, R.~Carballo-Rubio, and L.~J. Garay, ``{Black holes turn white
  fast, otherwise stay black: no half measures},'' JHEP {\bf 01} (2016) 157,
\href{http://arXiv.org/abs/1511.00633}{{\tt arXiv:1511.00633}}.

\bibitem{Barrau:2015uca}
A.~Barrau, B.~Bolliet, F.~Vidotto and C.~Weimer, ``{Phenomenology of bouncing
  black holes in quantum gravity: a closer look},'' JCAP \textbf{02} (2016) 022,
\href{http://arXiv.org/abs/1507.05424}{{\tt arXiv:1507.05424}}.

\bibitem{Vidotto:2018wvr}
F.~Vidotto, ``{Measuring the last burst of non-singular black holes},'' Found.
  Phys. \textbf{48} (2018) 1380,
\href{http://arXiv.org/abs/1803.02755}{{\tt arXiv:1803.02755}}.

\bibitem{Carballo-Rubio:2018jzw}
R.~Carballo-Rubio, F.~Di Filippo, S.~Liberati and M.~Visser, ``{Phenomenological
  aspects of black holes beyond general relativity},'' Phys. Rev. D \textbf{98}
  (2018) 124009,
\href{http://arXiv.org/abs/1809.08238}{{\tt arXiv:1809.08238}}.


\end{thebibliography}
\end{document}